\documentclass[conference]{IEEEtran}
\usepackage{cite}
\usepackage{amsmath,amssymb,amsfonts,amsthm}
\usepackage{stmaryrd}
\usepackage{algorithmic}
\usepackage{graphicx}
\usepackage{textcomp}
\usepackage{xcolor}
\usepackage{makecell}
\usepackage{optidef}
\usepackage{algorithm}
\usepackage{algorithmic}
\usepackage{adjustbox}
\usepackage{flushend}

\usepackage{subcaption}
\usepackage[all=normal,paragraphs=tight,floats=normal,mathspacing=normal,wordspacing=normal,charwidths=normal,mathdisplays=normal,leading=normal]{savetrees}

\usepackage{multirow}
\usepackage{array}
\usepackage{arydshln}
\usepackage{makecell}

\usepackage{booktabs}

\usepackage{tabulary}

\newcommand{\tagphantom}{\adjustbox{phantom}{(\arabic{equation})} \notag}
\newcommand\scalemath[2]{\scalebox{#1}{\mbox{\ensuremath{\displaystyle #2}}}}

\usepackage{array}

\makeatletter
\newcommand{\thickhline}{%
    \noalign {\ifnum 0=`}\fi \hrule height 1pt
    \futurelet \reserved@a \@xhline
}

\DeclareFontFamily{U}{wncy}{}
\DeclareFontShape{U}{wncy}{m}{n}{<->wncyr10}{}
\DeclareSymbolFont{mcy}{U}{wncy}{m}{n}
\DeclareMathSymbol{\Sh}{\mathord}{mcy}{"58} 

\newcommand{\abs}[1]{\ensuremath{\left| #1 \right|}}
\newcommand{\esp}[1]{\ensuremath{\mathbb{E}\left[ #1 \right]}}
\newcommand{\norm}[2]{\ensuremath{\left\lVert #1 \right\rVert}_#2}
\DeclareMathOperator*{\argmax}{arg\,max}
\DeclareMathOperator*{\argmin}{arg\,min}
\DeclareMathOperator*{\sinc}{sinc}

\def\BibTeX{{\rm B\kern-.05em{\sc i\kern-.025em b}\kern-.08em
    T\kern-.1667em\lower.7ex\hbox{E}\kern-.125emX}}

\SetSymbolFont{stmry}{bold}{U}{stmry}{m}{n}

\begin{document}

\title{Efficient Deep Unfolding for SISO-OFDM \\ Channel Estimation}

\author{
    \IEEEauthorblockN{Baptiste Chatelier\IEEEauthorrefmark{2}\IEEEauthorrefmark{3}, Luc Le Magoarou\IEEEauthorrefmark{2}\IEEEauthorrefmark{3}, Getachew Redieteab\IEEEauthorrefmark{4}\IEEEauthorrefmark{3}}
    \IEEEauthorblockA{\IEEEauthorrefmark{2}Univ Rennes, INSA Rennes, CNRS, IETR-UMR 6164, Rennes, France}
    \IEEEauthorblockA{\IEEEauthorrefmark{3}b\raisebox{0.2mm}{\scalebox{0.7}{\textbf{$<>$}}}com, Rennes, France}
    \IEEEauthorblockA{\IEEEauthorrefmark{4}Orange Innovation, Rennes, France}
    \IEEEauthorblockA{baptiste.chatelier@insa-rennes.fr}
}

\maketitle

\begin{abstract}
    In modern communication systems, channel state information is of paramount importance to achieve capacity. It is then crucial to accurately estimate the channel. It is possible to perform SISO-OFDM channel estimation using sparse recovery techniques. However, this approach relies on the use of a physical wave propagation model to build a dictionary, which requires perfect knowledge of the system's parameters. In this paper, an unfolded neural network is used to lighten this constraint. Its architecture, based on a sparse recovery algorithm, allows SISO-OFDM channel estimation even if the system's parameters are not perfectly known. Indeed, its unsupervised online learning allows to learn the system's imperfections in order to enhance the estimation performance. The practicality of the proposed method is improved with respect to the state of the art in two aspects: constrained dictionaries are introduced in order to reduce sample complexity and hierarchical search within dictionaries is proposed in order to reduce time complexity. Finally, the performance of the proposed unfolded network is evaluated and compared to several baselines using realistic channel data, showing the great potential of the approach.
\end{abstract}


\section{Introduction}
    Recently, machine learning techniques have emerged as a promising solution in many wireless communications areas~\cite{OShea2017,Wang2017} such as beamforming~\cite{Lemagoarou2022,Alkhateeb2018}, channel charting~\cite{Yassine2022b,Studer2018,Ferrand2021} or channel mapping~\cite{LeMagoarou2021a,Alrabeiah2019}. The Deep Unfolding machine learning approach~\cite{Shlezinger2020,Monga2021,Hershey2014,Balatsoukas2019} is very promising as it profits from both the controlled complexity of classical signal processing approaches and the flexibility of machine learning techniques. Indeed, this approach considers that iterative algorithms can be unfolded as neural networks, that can be trained, where each layer represents one iteration of the algorithm.

    In the field of channel estimation, classical statistical estimation techniques such as the Least Squares (LS) or the Minimum Mean Squared Error (MMSE) have been extensively used in the past. However, those techniques suffer from several drawbacks: on one hand, for the LS estimator, there is a huge Mean Squared Error (MSE), on the other hand, for the MMSE estimator, there is a huge computational complexity. Other approaches have been envisioned to counter those drawbacks: one of them is the usage of \textit{sparse recovery} algorithms. It is well known that propagation channels are dominated by a few strong propagation paths: the channel is then said to be \textit{sparse}. It has been shown that one can use that sparsity notion to propose channel estimation strategies relying on sparse recovery algorithms such as Matching Pursuit (MP)~\cite{Kang2008,Karabulut2004,Khojastepour2009,Yassine2022}. However, those techniques rely on the knowledge of a physical wave propagation model, and it has been shown in~\cite{Yassine2022} that a small uncertainty on system parameters could lead to high estimation performance losses. 

    \noindent \textbf{Contributions and related work.}
    In~\cite{Yassine2022}, the Deep Unfolding approach was used to solve this performance loss issue, with the unfolded network mpNet, allowing to learn the physical imperfections of the system so as to enhance the estimation performance. However, this approach yields a high number of parameters to learn, leading to both a high sample complexity and a high time complexity. In this paper, mpNet is adapted to constrained dictionaries, which allows to drastically reduce sample complexity. Moreover, hierarchical search over dictionaries is introduced in order to reduce time complexity. Furthermore, the proposed method is illustrated on a Single Input Single Output Orthogonal Frequency Division Multiplexing (SISO-OFDM) system here instead of a Multi-User Multiple Input Multiple Output (MU-MIMO) system as in~\cite{Yassine2022}. However, note that the contributions apply equally to both types of systems.

    The rest of the paper is organized as follows, Section~\ref{2} presents the SISO-OFDM channel estimation problem, the used physical wave propagation model, and the sparse recovery approach. Section~\ref{3} presents the unfolded neural network and the two contributions of this paper: the sample complexity reduction strategy, and the time complexity reduction strategy. Section~\ref{4} presents simulation results. Finally, Section~\ref{Conclusion} concludes the paper and gives perspectives. 


\section{Problem formulation}\label{2}
    \subsection{System model}
    In this paper, a SISO-OFDM scenario with $N$ subcarriers in the uplink is considered. Let $\mathbf{g} \in \mathbb{C}^N$ denote the antenna gain vector representing the Base Station (BS) antenna gain at each subcarrier frequency, and $\mathbf{f} \in \mathbb{R}^N$ denote a frequency vector containing the different subcarrier frequencies. Finally, a multipath channel with $L_p$ propagation paths is considered.

    Let $\mathbf{x} \in  \mathbb{C}^N$ be an LS channel estimate of the channel vector $\mathbf{h} \in  \mathbb{C}^N$. This estimate is noisy as it possesses a residual estimation error:
    \begin{equation}
        \mathbf{x} = \mathbf{h}+\mathbf{n},
    \end{equation}
    where $\mathbf{n}\sim \mathcal{CN}\left(0,\sigma^2 \mathbf{I}_N\right)$. Its Signal to Noise Ratio (SNR) can be computed as:
    \begin{equation}
        \text{SNR}_{\text{in}} = \dfrac{\norm{\mathbf{h}}{2}^2}{N \sigma^2}.
    \end{equation}

    The goal of this paper is to denoise this LS channel estimate using sparse recovery techniques and a physical model.
    
    \subsection{Physical model}
    For a given subcarrier $f_k$, the channel coefficient can be written as:
    \begin{equation}
        h_k = \sum_{l=1}^{L_p} \alpha_l g_k e^{-\mathrm{j}2\pi f_k \tau_l},
    \end{equation}
    where $\alpha_l$ and $\tau_l$ respectively represent the complex attenuation coefficient and the propagation delay for the $l^{\text{th}}$ path.
    
    Under this model, the channel vector $\mathbf{h} \in \mathbb{C}^N$ can be expressed as:
    \begin{equation}
        \mathbf{h} = \sum_{l=1}^{L_p} \alpha_l \begin{bmatrix}
            g_1 e^{-\mathrm{j}2\pi f_1 \tau_l}\\
            \vdots \\
            g_N e^{-\mathrm{j}2\pi f_N \tau_l}
        \end{bmatrix} = \sum_{l=1}^{L_p} \alpha_l \boldsymbol{\psi}\left(\tau_l\right),
    \end{equation}
    where $\boldsymbol{\psi}\left(\tau_l\right)$ is the \textit{frequency response vector} (FRV) for the $l^{\text{th}}$ path. We can see that the channel vector can be defined as a linear combination of FRVs. The linear combination is said to be sparse when $L_p$ is small.

    \subsection{Sparse recovery approach}
    As the channel can be written as a linear combination of FRVs, it is possible to construct a fixed matrix $\mathbf{\Psi} \in \mathbb{C}^{N\times A}$ with $A$ FRV columns, which amounts to the discretization of the delays, and a projection vector $\mathbf{u} \in \mathbb{C}^A$ such that the channel could be estimated as:
    \begin{equation}
        \hat{\mathbf{h}} = \mathbf{\Psi u}.
    \end{equation}
    For the rest of this paper, the FRV matrix $\mathbf{\Psi}$ will be called \textit{dictionary} and its columns will be called \textit{atoms}, i.e. each atom will represent a FRV corresponding to a specific delay. As the channel vector can be seen as a linear combination of a few FRVs, the channel vector is said to be sparse in a dictionary of FRVs. For a fixed dictionary $\mathbf{\Psi}$, the sparse recovery optimization problem is:
    \begin{mini}|s|
        {\mathbf{u}}{\norm{\mathbf{h}-\mathbf{\Psi u}}{2}^2}{}{}
        \addConstraint{\norm{\mathbf{u}}{0}\leq A}
        \label{eq:optim}
    \end{mini}
    The previous optimization problem makes sense only if the dictionary is correct, i.e. a sparse linear combination of dictionary columns can represent a channel vector. If this is not the case, the channel estimate won't be trustworthy. It is possible to find the optimal dictionary as:  
    \begin{equation}
        \mathbf{\Psi}^{\star} = \argmin_{\mathbf{\Psi}} \mathbb{E}_{\mathbf{h}}\left[\min_{\mathbf{u}, \norm{\mathbf{u}}{0}\leq A}\norm{\mathbf{h}-\mathbf{\Psi u}}{2}^2\right].
    \end{equation}
    Solving Eq.~\eqref{eq:optim} for a fixed FRV dictionary is possible through sparse recovery algorithm such as the MP algorithm~\cite{Mallat1993}: this is the approach followed in this paper, as in~\cite{Yassine2022}. The goal of this algorithm is to find the most correlated FRVs, i.e. atoms, of the dictionary with the noisy channel estimate and subtract their projections. After $L_p$ iterations of the algorithm, in a perfect scenario, as all the corresponding FRVs will have been subtracted to the residual, the residual will only be composed of noise.


    \begin{algorithm}
        \caption{MP algorithm for channel denoising}
        \label{Algo2}
            \begin{algorithmic}[1]
                \REQUIRE Dictionary $\mathbf{\Psi}$, Noisy LS estimate $\mathbf{x}$, Tolerance level $\epsilon$
                \STATE Initialize the residual: $\mathbf{r} \leftarrow \mathbf{x}$
                \WHILE{$\norm{\mathbf{r}}{2}^2>\epsilon$}
                    \STATE Find the most correlated atom: $i^{\star} \leftarrow \argmax_{i} \abs{\boldsymbol{\psi}_i^H \mathbf{r}}$
                    \STATE Update the residual: $\mathbf{r}\leftarrow \mathbf{r}- \boldsymbol{\psi}_{i^{\star}} \boldsymbol{\psi}_{i^{\star}}^H \mathbf{r}$
                \ENDWHILE
                \ENSURE $\hat{\mathbf{h}} \leftarrow \mathbf{x} - \mathbf{r}$ (Denoised LS estimate)
            \end{algorithmic}
    \end{algorithm}
    
    \subsection{Hardware impairments}

    In reality, the FRVs are not exactly known as there exists a lot of hardware imperfections related to frequency generation/acquisition. More specifically, Carrier Frequency Offset (CFO) and Sampling Clock Offset (SCO) can offset the subcarrier frequencies. Let $\tilde{f_i}$ be the \textit{nominal} $i^{\text{th}}$ subcarrier frequency, i.e. the non-offseted subcarrier frequency, and $f_i$ be the \textit{real} $i^{\text{th}}$ subcarrier frequency, i.e. the potentially offseted subcarrier frequency. For the CFO, the frequency offset $\delta f$ is common to all subcarriers:
    \begin{equation}
        \forall i \in \Biggl\llbracket -\dfrac{N}{2}, \dfrac{N}{2} \Biggr\rrbracket, \text{ } f_i = \tilde{f}_i + \delta f.
    \end{equation}
    For the SCO, it has been shown in~\cite{Kim1998,Speth1999} that the frequency offset is dependent on the subcarrier index, on the oscillator part per million (ppm) value $\xi$, and on the subcarrier spacing $\Delta f$:
    \begin{equation}
        \forall i \in \Biggl\llbracket -\dfrac{N}{2}, \dfrac{N}{2} \Biggr\rrbracket, \text{ } f_i = \tilde{f}_i + i \delta f = \tilde{f}_i + i \xi \Delta f.
    \end{equation}
    Moreover, we consider an antenna gain imperfection. Let $\tilde{g_i}$ be the nominal antenna gain, and $g_i$ be the real antenna gain for the $i^{\text{th}}$ subcarrier. We obtain:
    \begin{equation}
        \forall i \in \Biggl\llbracket -\dfrac{N}{2}, \dfrac{N}{2} \Biggr\rrbracket, \text{ } g_i = \tilde{g}_i + n_{g_{i}},
    \end{equation}
    with $n_{g_{i}} \sim \mathcal{N}\left(0,\sigma^2_g\right)$.

    For the rest of this paper, we will only consider SCO and antenna gain imperfections. 
    We can then define the notions of \textit{nominal} and \textit{real dictionaries}. The nominal dictionary $\tilde{\mathbf{\Psi}}$ will represent the dictionary constructed from the nominal knowledge of the system parameters, which is unaware of impairments. On the other hand, the real dictionary will represent the dictionary constructed from the perfect knowledge of the system parameters, i.e. with full knowledge of the SCO and antenna gain imperfections. The proposed method consists in initializing a neural network using the nominal dictionary, and approaching the performance one would get using the real dictionary via learning.

\section{Proposed approach}\label{3}

    \subsection{mpNet architecture}
    Using the same approach as in~\cite{Yassine2022}, we propose to use the unfolded neural network mpNet to achieve SISO-OFDM channel estimation while learning the physical imperfections. The network is the unfolded version of the MP iterative algorithm and is presented in Algorithm~\ref{Algo_mpNet}. 
    
    \begin{algorithm}
        \caption{Forward pass of mpNet~\cite{Yassine2022}}
        \label{Algo_mpNet}
            \begin{algorithmic}[1]
                \REQUIRE Weight matrix $\mathbf{W}\in \mathbb{C}^{N\times A}$, Noisy LS estimate $\mathbf{x} \in \mathbb{C}^N$, Noise variance $\sigma^2$, Number of subcarriers $N$
                \STATE $\mathbf{r} \leftarrow \mathbf{x}$
                \STATE $\epsilon \leftarrow \frac{\sigma^2N}{\norm{\mathbf{x}}{2}^2}$
                \WHILE{$\norm{\mathbf{r}}{2}^2>\epsilon$}
                    \STATE $\mathbf{r} \leftarrow \mathbf{r} - \mathbf{W}\text{HT}_1\left(\mathbf{W}^H \mathbf{r}\right)$
                \ENDWHILE
                \ENSURE $\hat{\mathbf{h}} \leftarrow \mathbf{x} - \mathbf{r}$ (Denoised LS estimate)
            \end{algorithmic}
    \end{algorithm}
    $\text{HT}_1$ is a non-linear hard-thresholding operator which keeps only its input of greatest modulus and sets all the others to zero. The weight matrix $\mathbf{W}$ is initialized with the nominal dictionary $\tilde{\mathbf{\Psi}}$. This thoughtful initialization allows to reduce the cost function convergence time in comparison to a random initialization. The goal is to learn the optimal dictionary through gradient descent in the unfolded network. In brief, this network has two goals: 
    \begin{itemize}
        \item The forward pass of mpNet performs the channel estimation/denoising operation as the architecture of mpNet is the unfolded MP algorithm. Each layer will estimate one path of the channel.
        \item The backward pass of mpNet performs the dictionary learning. As the weight matrix of mpNet is initialized with the nominal dictionary, the goal of the backward pass is to update the weight matrix so that it tends towards the real dictionary.
    \end{itemize}

    Moreover, the value $\epsilon$ in Algorithm~\ref{Algo_mpNet} corresponds to a stopping criterion called  $\mathtt{SC_2}$ in~\cite{Yassine2022}:
    \begin{equation}
        \mathtt{SC_2} : \norm{\mathbf{r}}{2}^2 \leq \dfrac{\sigma^2}{\norm{\mathbf{x}}{2}^2}N.
    \end{equation}
    This stopping criterion allows the network to have an adaptive number of layers for given noise characteristics: in other words, SNR adaptability. However, it imposes the knowledge of the noise variance $\sigma^2$ at the BS side.

    \subsection{Reducing sample complexity: constrained dictionary}\label{C_mpNet}

    As we explained above, the learning parameters are the components of the weight matrix $\mathbf{W} \in \mathbb{C}^{N\times A}$. As those components are complex, each entry of $\mathbf{W}$ possesses both a magnitude and a phase: as a result there are $2NA$ learning parameters. In the experimental setup that will be presented in Section~\ref{4}, we will consider $N=256$ subcarriers, and $A=990$ atoms. In that setup, there are $2NA = 506,880$ learning parameters. The first important contribution of this paper is to reduce this number by using a \textit{constrained} dictionary.

    We know that the proposed SISO-OFDM system only considered imperfections that impacted the gains and frequencies of the dictionary. Moreover, gain and frequencies are expected to be independent of the propagation delays, meaning that they will be common for each atom of the dictionary. We can then propose the following definition for the constrained dictionary $\overset{\diamond}{\mathbf{\Psi}} \in \mathbb{C}^{N\times A}$:
    \begin{equation}
        \overset{\diamond}{\mathbf{\Psi}} =  \scalemath{0.93}{\begin{bmatrix}
            g_1e^{-\mathrm{j}2\pi \left(f_1-\frac{N}{2}\delta f\right) \tau _1} & \cdots & g_1e^{-\mathrm{j}2\pi \left(f_1-\frac{N}{2}\delta f\right) \tau _A} \\
            \vdots & \vdots & \vdots \\
            g_Ne^{-\mathrm{j}2\pi \left(f_N+\frac{N}{2}\delta f\right) \tau _1} & \cdots & g_Ne^{-\mathrm{j}2\pi \left(f_N+\frac{N}{2}\delta f\right) \tau _A}
        \end{bmatrix}}
    \end{equation}
    Each column of the dictionary still represents an FRV, however each entry of this constrained matrix is no more learnable: the only learnable parameters are the complex antenna gains $\mathbf{g}$ and the SCO frequency offset $\delta f$. From $2NA$ parameters, we only have $2N+1$ learning parameters left. This is interesting as \emph{the number of learning parameters is now independent of the number of atoms}. If we take the previous numerical example, we go from $2NA=506,880$ to only $2N+1=513$ learning parameters.

    One of the other benefits of this constrained dictionary approach is its sample complexity, i.e. the number of required samples to achieve convergence of the network's cost function. Indeed, for the classical mpNet architecture, as each component of the weight matrix was independent, only the atoms selected in the forward pass were updated during backpropagation. For the proposed approach, as the frequency offset and gains are common over all atoms, the selection and update of one atom will result in the update of all other atoms. Consequently, the cost function convergence will be quicker with this constrained dictionary approach than with the classical mpNet architecture, as shown in the experimental part of this paper.

    \subsection{Reducing time complexity: hierarchical search}\label{section_argmin}

    A quick complexity analysis allows us to state that the time complexity of the classical mpNet approach is $\mathcal{O}\left(DNA\right)$ where $D$ represents the number of layers of the mpNet network. If we take the previously considered scenario with $D=10$ layers, we obtain at least $DNA = 2,534,400$ arithmetic operations for the estimation of one SISO-OFDM channel. In order to reduce this time complexity, we propose a new unfolded network architecture based on hierarchical atom search. 

    \begin{figure}[!h]
        \centering
        \includegraphics[scale=.8]{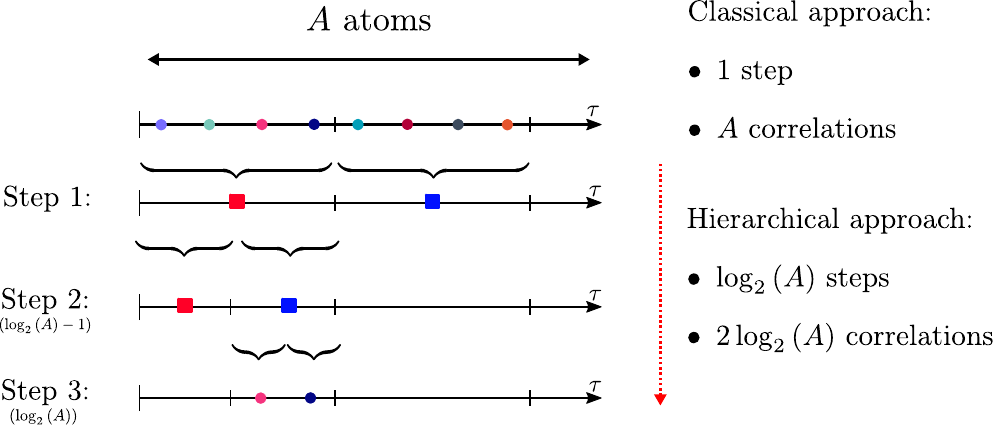}
        \caption{Hierarchical atom search}
        \label{fig:hierarch_search}
    \end{figure}

    The principle of the hierarchical atom search approach is presented in Fig.~\ref{fig:hierarch_search}. The classical approach consists in finding the most correlated atom in a dictionary of $A$ different atoms (represented as circles in Fig.~\ref{fig:hierarch_search}): $A$ different correlations are carried out. Our hierarchical approach consists in iteratively finding the most correlated atom in the dictionary. As in the SISO-OFDM scenario the atoms are parameterized with a propagation delay, the dictionary covers an interval in the delay domain. In the first step, we propose to construct two \textit{meta-atoms} (represented as squares in Fig.~\ref{fig:hierarch_search}) whose correlation responses separately cover the dictionary-associated delay interval. We carry out the correlation with the two meta-atoms and choose the most correlated one. We then create two new meta-atoms that cover the delay interval associated with the most correlated meta-atom of the previous step. This process is repeated until the $\log_2 \left(A\right)^{\text{th}}$ step. In the delay interval covered by the most correlated atom chosen in the $\log_2 \left(A\right)-1^{\text{th}}$ step, there are only $2$ atoms from the original dictionary: we compute their correlations and then find the most correlated atom of the dictionary. In other words, rather than finding the most correlated atom in $1$ step but with $A$ correlations, we find the most correlated atom in $\log_2\left(A\right)$ steps but with only $2\log_2\left(A\right)$ correlations.

    The challenge of this approach is to propose a meta-atom structure that allows its correlation response to cover a specific delay interval. We can prove that using a sinc modulated FRV as a meta-atom allows this. 
    
\begin{figure}[!h]
        \centering
        \includegraphics[scale=.35]{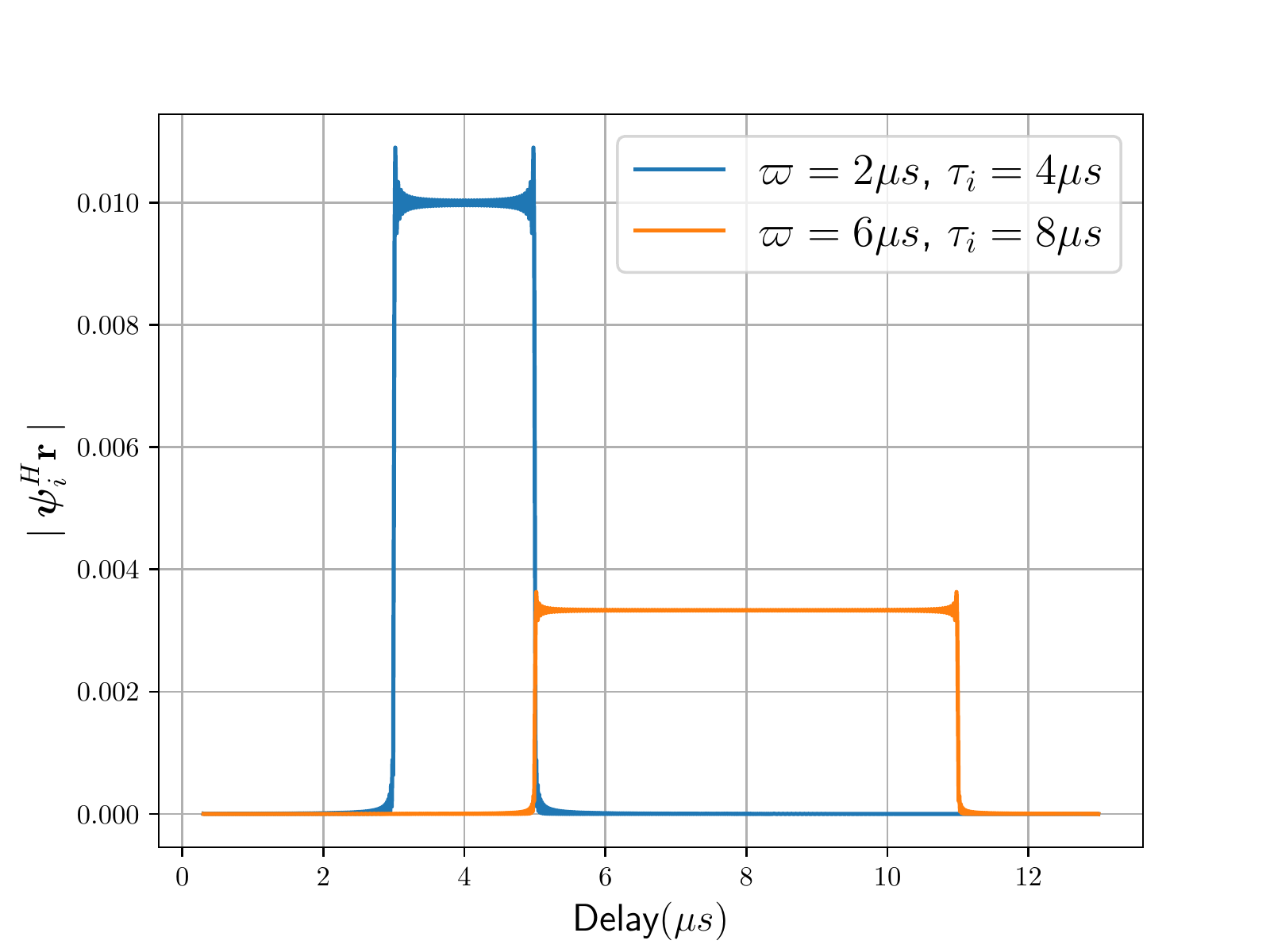}
        \caption{Influence of $\varpi$ and $\tau_i$ on the correlation waveform}
        \label{fig:hierarch_waveform}
    \end{figure}

    A Fourier Transform $\text{FT}$ from the frequency domain to the delay domain is defined as:
    \begin{equation}
        x\left(\tau\right) = \text{FT}\left[x\left(f\right)\right] = \int_{\mathbb{R}} x\left(f\right)e^{-\mathrm{j}2\pi f\tau}\mathrm{d}f.
    \end{equation}
    The correlation of a pure FRV of delay $\tau$, $e\left(f,\tau\right)$, and a generic meta-atom $\psi_i\left(f\right)$ is considered. The bandwidth, central frequency, and subcarrier spacing of the system will respectively be denoted as $\gamma$, $f_0$, and $\Delta f$. The correlation response can be computed as:
    \begin{align}
        \chi_{\psi_i} \left(\tau\right) &= \langle e\left(f,\tau\right), \psi_i\left(f\right)\rangle\\
        &= \int_{\mathbb{R}} \psi_i^*\left(f\right) g\left(f\right)e^{-\mathrm{j}2\pi f\tau } \Pi_{\gamma}\left(f-f_0\right) \Sh_{\Delta f} \left(f\right) \mathrm{d}f.
    \end{align}
    The windowing $\Pi_{\gamma}\left(f-f_0\right)$ allows to represent the band-limited nature of the FRV and meta-atom, while the Dirac comb $\Sh_{\Delta f} \left(f\right)$ allows the sampling at the subcarrier frequencies. We make the assumption that the antenna gains are quasi-constant over the bandwidth so that $g\left(f\right) = \alpha \Pi_\gamma \left(f-f_0\right)$. In order to simplify the following equations, we will deliberately omit the constant $\alpha$. We then obtain:
    \begin{align}
        \chi_{\psi_i} \left(\tau\right) &= \text{FT}\left[\psi_i^*\left(f\right)\Pi_{\gamma}\left(f-f_0\right) \Sh_{\Delta f} \left(f\right)\right]. \label{eq:ft_corr}
    \end{align}
    One can observe in Eq.~\eqref{eq:ft_corr} that the correlation response can be computed as a Fourier Transform. As we want it to be a constant over a certain delay interval (a rect function), it is then easy to see that using sinc modulated FRVs allows to fulfill that requirement. More specifically, we define the meta-atom $\psi_i \left(f\right)$ as:
    \begin{equation}
        \psi_i \left(f\right) = \sinc\left(\varpi f\right)e^{-\mathrm{j}2\pi f \tau_i}.
    \end{equation}
    The correlation response then becomes:
    \begin{align}
        \chi_{\psi_i} \left(\tau\right) &=  \text{FT}\left[\sinc\left(\varpi f\right)e^{\mathrm{j}2\pi f \tau_i}\Pi_{\gamma}\left(f-f_0\right) \Sh_{\Delta f} \left(f\right)\right] \tagphantom \\
        &=  \scalemath{0.93}{\tfrac{1}{\varpi} \Pi_{\varpi}\left(\tau-\tau_i\right) \circledast \gamma \sinc\left(\gamma \tau\right) e^{-\mathrm{j}2\pi f_0 \tau} \circledast \tfrac{1}{\Delta f} \Sh_{\tfrac{1}{\Delta f}}\left(f\right)}.\label{eq:meta_atom_corr}
    \end{align}
    One can see in Eq.~\eqref{eq:meta_atom_corr} that the proposed meta-atom expression allows to obtain a rectangular waveform in the delay domain, $\varpi$ allowing the rect width control and $\tau_i$ the rect central position control. This can be seen in Fig.~\ref{fig:hierarch_waveform} where we vary the values of $\varpi$ and $\tau_i$.

    For two meta-atoms, the time complexity of the proposed hierarchical search can be evaluated as $\mathcal{O}\left(2\log_2\left(A\right)\right)$. If we consider the partition of $A>1$ atoms into $n\in \mathbb{N}, \text{ } n>1$ intervals, the time complexity of the associated hierarchical search is $\mathcal{O}\left(n\log_n\left(A\right)\right)$. Minimizing the time complexity of this hierarchical approach is then equivalent to solving Eq.~\eqref{eq:min}:
    \begin{equation}
        n^\star = \argmin_{n\in \mathbb{N}, \text{ } n>1} n\log_n \left(A\right).
        \label{eq:min}
    \end{equation}
    It can be easily shown that solving Eq.~\eqref{eq:min} gives us $n^\star =3$.

    A quick complexity analysis of all the unfolded networks presented so far is presented in Table~\ref{table:complex}. We compare the classical mpNet as it was proposed in~\cite{Yassine2022}, mpNet with a constrained dictionary (C. mpNet), and mpNet with a constrained dictionary and hierarchical atom search with $3$ meta-atoms (H.C.$3$ mpNet). As we stated earlier, the constrained dictionary approach allows to reduce the sample complexity and the hierarchical search approach was developed to reduce the time complexity. It is then interesting to merge the constrained and hierarchical search approach, as it allows to reduce both the sample and time complexities. For $D=10$ layers, $N=256$ subcarriers and $A=990$ atoms, the time complexity of the forward pass goes from $2,534,400$ arithmetic operations for the classical mpNet to around only $48,220$ arithmetic operations for the H.C.$3$ mpNet, which is a fifty-fold reduction.

    \begin{table}
        \centering
        \scriptsize
            \begin{tabular}{cccccc}
                \toprule
                & & mpNet & C. mpNet & H.C.$3$ mpNet \\ \midrule
                \multirow[c]{2}{*}{\vspace{-2mm} \makecell{Time \\ complexity }}& Forward & $\mathcal{O}\left(DNA\right)$ & $\mathcal{O}\left(DNA\right)$ & $\mathcal{O}\left(DN3\log_3\left(A\right)\right)$\\
                \cmidrule(l){2-6}& Backward& $\mathcal{O}\left(DN\right)$ & $\mathcal{O}\left(DN\right)$ & $\mathcal{O}\left(DN\right)$\\
                \midrule
                \makecell{Sample \\ complexity } & &  +++ & + & +\\
                \bottomrule
            \end{tabular}
            \caption{Unfolded network complexity analysis}
            \label{table:complex}
    \end{table}

\section{Experimental results}\label{4}
    We propose to validate our contributions using realistic channel data, namely the ``O1" outdoor DeepMIMO dataset~\cite{Alkhateeb2019}. We consider a central frequency of $3.4$GHz, and a bandwidth of $50$MHz. Furthermore, we generate the channels with $N=256$ subcarriers. The dictionary is composed of $A=990$ atoms. We consider up to $L_p=10$ propagation paths. Regarding the physical imperfections, the oscillator ppm value for the SCO is $\xi=40$ppm, and we consider two gain noise variance scenarios: $\sigma^2_g=0.36$ (very high gain noise) and $\sigma^2_g=0.09$ (high gain noise). We consider two SNR scenarios: $\text{SNR}_{\text{in}}=10$dB and $\text{SNR}_{\text{in}}=5$dB. In order to compute the stopping criterion for the network, we assume that the noise variance is known at the BS side. 

\begin{figure}[htbp]
        \centering
        \includegraphics[scale=.4]{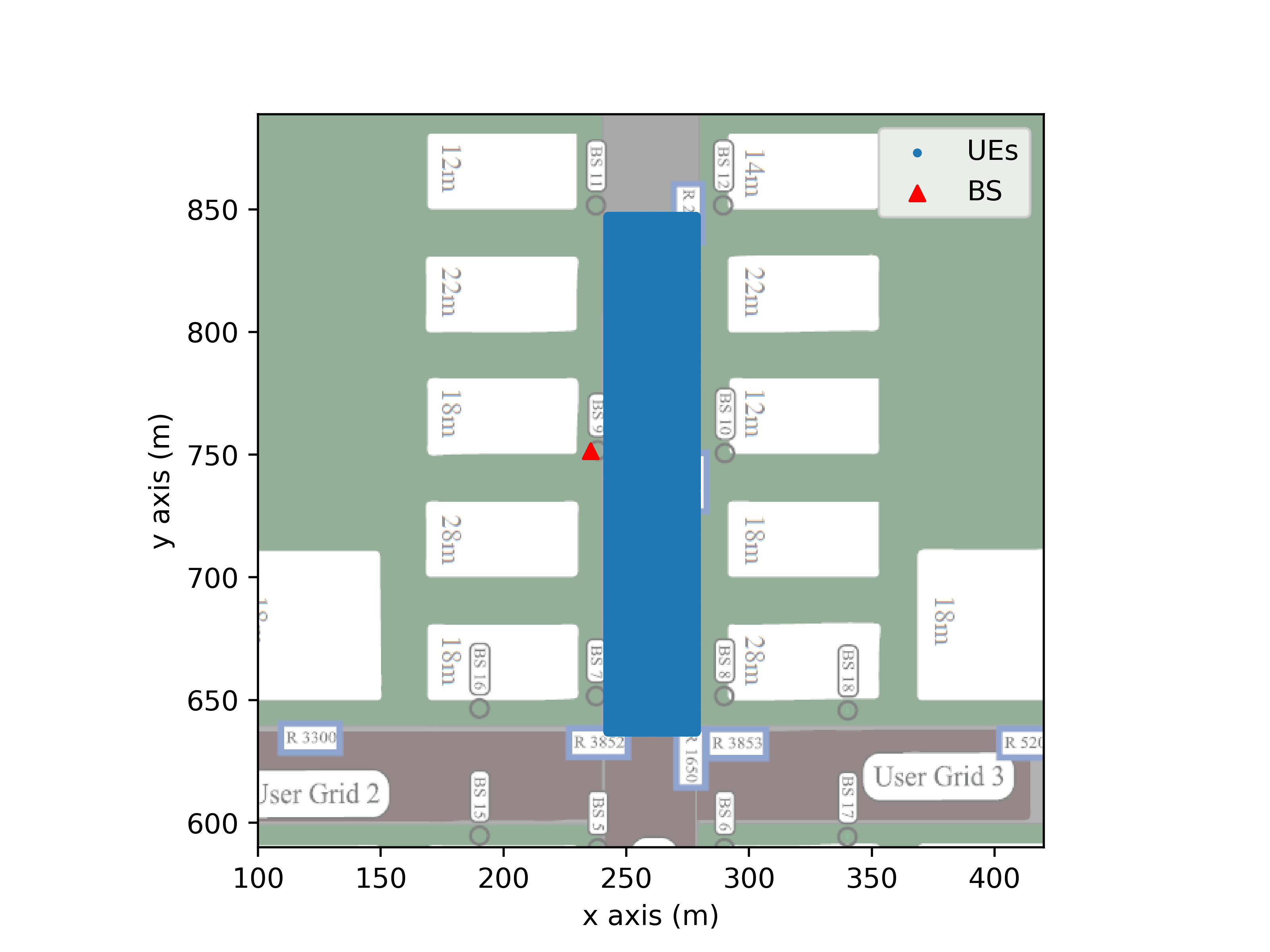}
        \caption{DeepMIMO channel generation configuration}
        \label{fig:DeepMIMO}
    \end{figure}

    We consider the same minibatch online learning strategy as in~\cite{Yassine2022}. The network will see batches of $10$ channels, and the estimation performance will be evaluated on $2000$ test channels. More specifically, regarding the DeepMIMO parameters, we consider the BS represented with a red triangle in Fig.~\ref{fig:DeepMIMO}, and User Equipments (UEs) randomly generated in the blue zone. We will use the Adam optimizer~\cite{Kingma2014} for the learning parameter update. The evaluation metric will be the Normalized Mean Squared Error (NMSE), defined as $\mathbb{E}\left[\lVert \hat{\mathbf{h}}-\mathbf{h} \rVert_2^2\right]/\lVert \mathbf{h} \rVert_2^2$.
    
    We propose to evaluate the performance of the classical network (mpNet), the constrained dictionary mpNet (C. mpNet), the hierarchical search with $2$ meta-atoms and constrained dictionary mpNet (H.C.$2$ mpNet) and the one with $3$ meta-atoms (H.C.$3$ mpNet). We will compare those network performances against the LS estimator, the MP algorithm with nominal and real dictionaries, and the MMSE estimator with a substituted channel covariance matrix using the Low-Rank Approximation (LRA) presented in~\cite{Savaux2017}. We use the substituted covariance matrix, as, in reality, estimating the covariance matrix for each UE is too complex as it requires tracking each UE channel during several frames.
    \begin{figure*}[t!]
        \captionsetup{font=small}
        \begin{subfigure}[b]{0.330\textwidth}
        \includegraphics[width=\columnwidth]{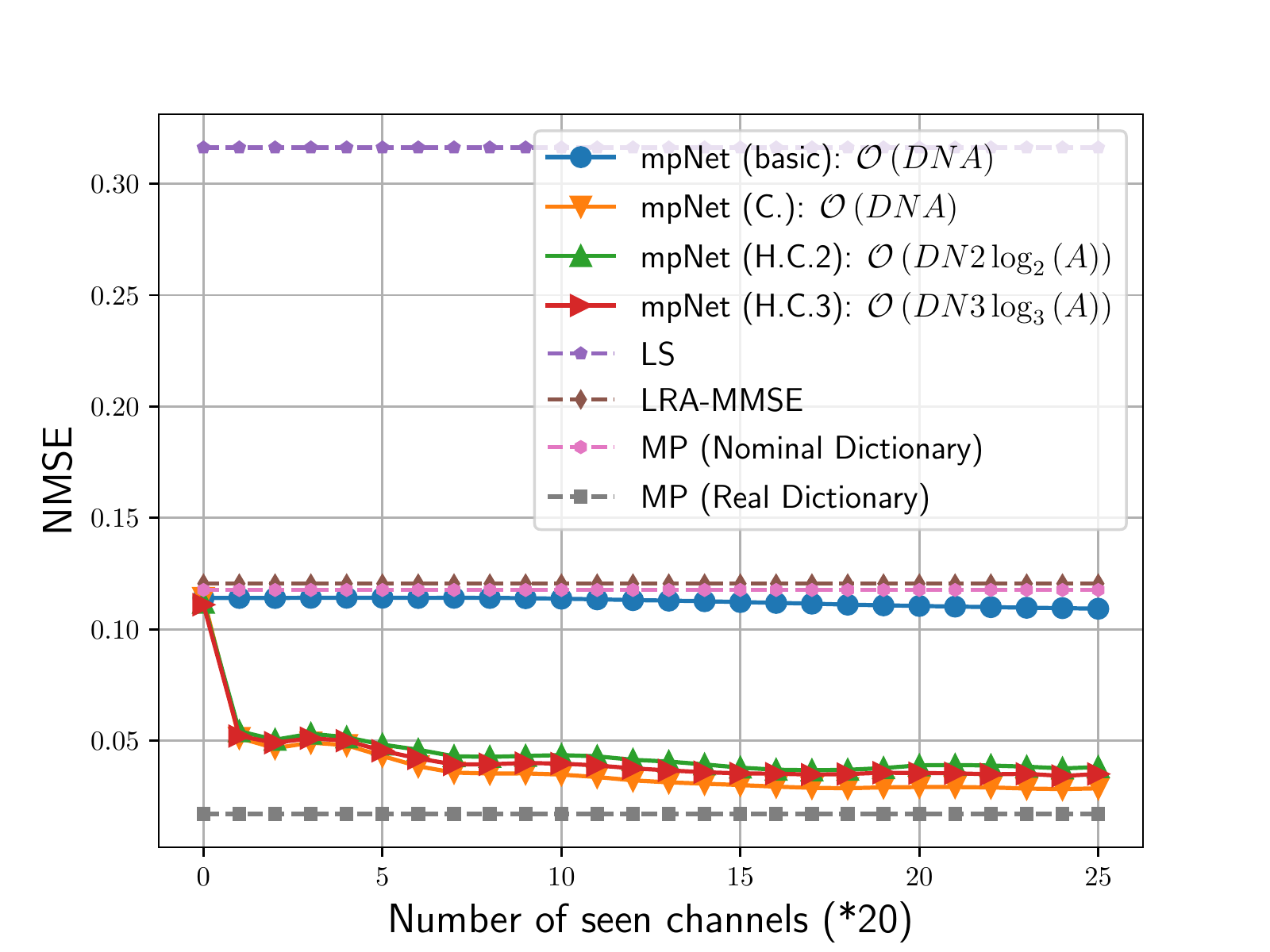}
        \caption{$\text{{SNR}}_{{\text{{in}}}}={5}\,\text{{dB}},\, \sigma^2_g={0.09},\, \xi=40\text{ ppm}$}\label{fig:multi_a}
        \end{subfigure}
        \begin{subfigure}[b]{0.330\textwidth}
        \includegraphics[width=\columnwidth]{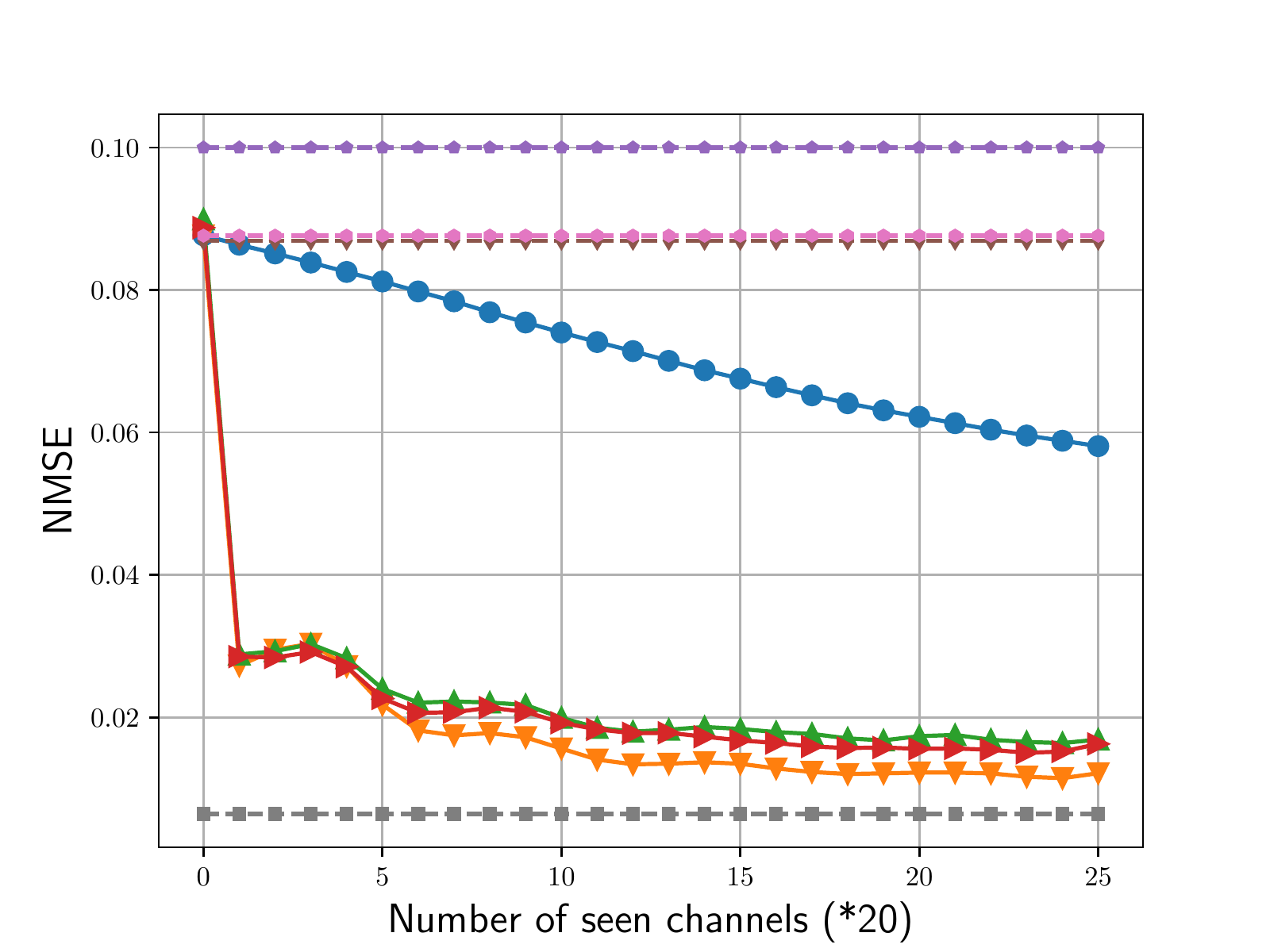}
        \caption{$\text{{SNR}}_{{\text{{in}}}}={10}\,\text{{dB}},\, \sigma^2_g={0.09},\, \xi=40\text{ ppm}$}\label{fig:multi_b}
        \end{subfigure}
        \begin{subfigure}[b]{0.330\textwidth}
        \includegraphics[width=\columnwidth]{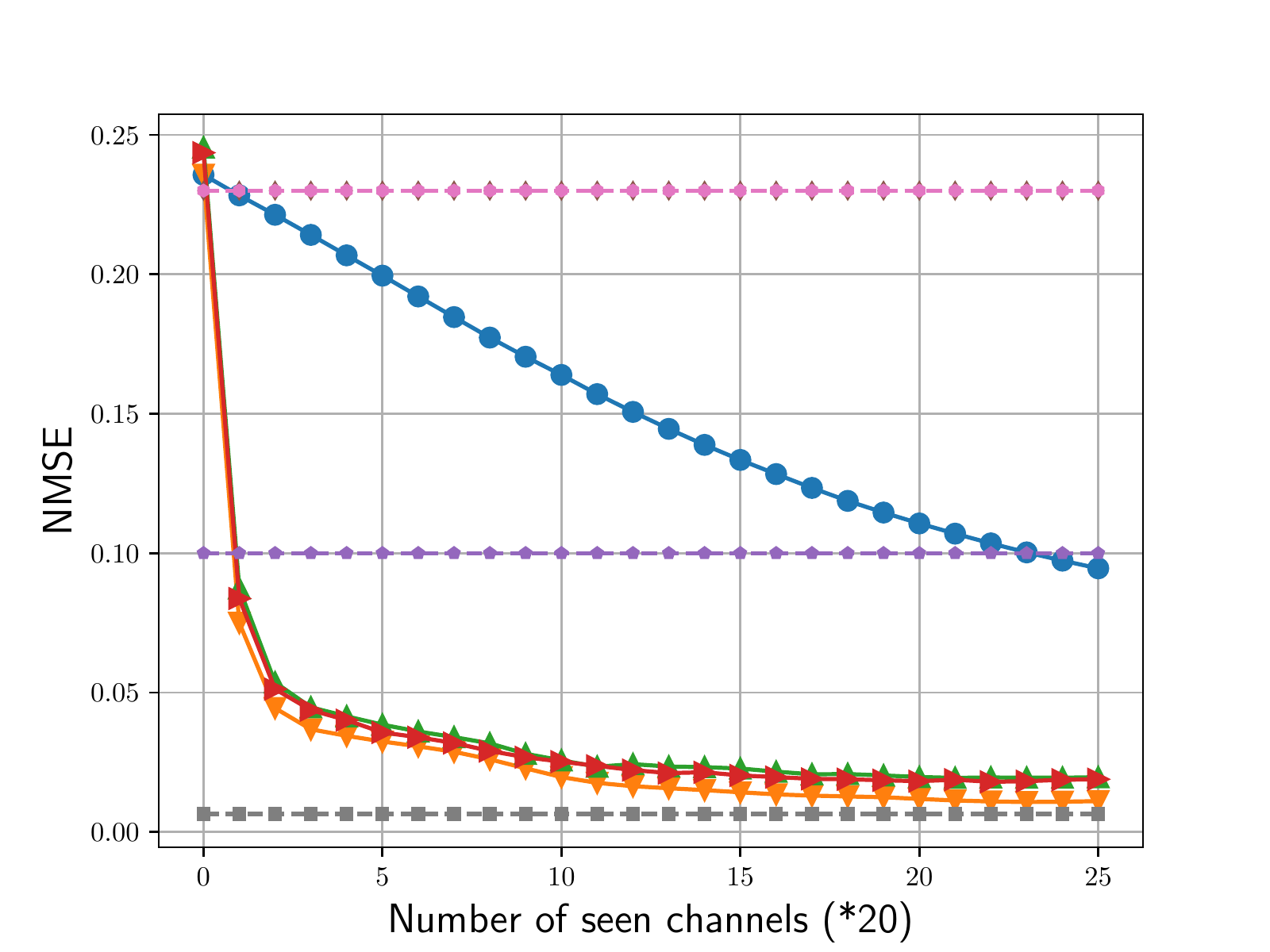}
        \caption{$\text{{SNR}}_{{\text{{in}}}}={10}\,\text{{dB}},\, \sigma^2_g={0.36},\, \xi=40\text{ ppm}$}\label{fig:multi_c}
        \end{subfigure}
        \caption{Channel estimation performance for various SNRs and model imperfections}
        \label{fig:multi}
    \end{figure*}
    
    For all noise and hardware scenarios, it is possible to remark in Fig.~\ref{fig:multi} that dictionary learning is achieved as there is convergence of the NMSE performances. One can remark that at the end of the training, the proposed networks (C. mpNet, H.C.$2$ and H.C.$3$ mpNet) always outperform the LS and LRA-MMSE approaches.
    
    One can see that, as exposed in Section~\ref{C_mpNet}, the convergence of the constrained versions of mpNet (C. mpNet and H.C. mpNet) is quicker than the convergence of the classical mpNet. For all scenarios, only approximately $400$ channels are needed to achieve convergence. Moreover, we can see that, after convergence, the estimation performances of the constrained networks are better than the one of the classical mpNet. This can be explained by the bias-variance decomposition of the estimation NMSE. In fact, the C./H.C. mpNet can be seen as an estimator with fewer parameters than the classical mpNet, so its variance will be lower than the classical mpNet one. As the dictionary structure is chosen wisely, the C./H.C. mpNet has the same bias as the classical mpNet. As a result, the estimation NMSE is lower for both the C. and H.C. mpNet architectures than for the classical one.

    One can also remark in Fig.~\ref{fig:multi_c} that, if the hardware imperfections are too important, the model-based MP algorithm does not perform well. In that particular case, using the MP approach with nominal dictionary is worse than using LS.

    It is noticeable that H.C.$2$ mpNet offer slightly worse performance than the C. mpNet. This can be explained by the hierarchical search approach: in the classical approach, the algorithm finds the most correlated atom of a dictionary with a residual. For the hierarchical approach, the algorithm will select the delay interval that has the most energy. As a result, for a fixed noisy LS estimate, the two approaches may not select the same atoms, leading to different channel estimation performance. It is then interesting to study the impact of the number of meta-atoms, as increasing the number of windows in the delay domain can improve the estimation performance for the hierarchical approach. This can be seen with the NMSE performance of H.C.$3$ mpNet: this network offers slightly better performance than the H.C.$2$ mpNet, while presenting a lower time complexity as shown in Table~\ref{table:complex}.

    It has been shown in Section~\ref{section_argmin} that the hierarchical approach with $3$ meta-atoms was optimal in terms of time complexity. This can be seen in Table~\ref{table:time_exec} where an execution time comparison of the MP algorithm to its hierarchical versions with $2$ and $3$ meta-atoms (MP H.$2$ and MP H.$3$) is proposed. One can see that the MP H.$3$ outperforms both the MP H.$2$ and classical MP algorithm: its mean execution time is up to a hundred time quicker than the classical MP.

    \begin{table}
        \centering
        \scriptsize
            \begin{tabular}{cccccc}
                \toprule
                & & MP & MP H.$2$  & MP H.$3$ \\ 
                \midrule
                \multirow[c]{3}{*}{\vspace{-2mm} $\esp{t_{\text{exec}}}$ (s)} & $A = 990$ & $0.048$ & $0.014$ & $\mathbf{0.013}$\\
                \cmidrule(l){2-6}& $A = 5,000$  & $0.263$ & $0.019$ & $\mathbf{0.015}$\\
                \cmidrule(l){2-6}& $A = 50,000$ & $2.866$ & $0.024$ & $\mathbf{0.020}$\\
                \bottomrule
            \end{tabular}
            \caption{Average execution time for $10,000$ random UEs}
            \label{table:time_exec}
    \end{table}

\section{Conclusion}\label{Conclusion}

    In this paper, the unfolded neural network mpNet \cite{Yassine2022} was enhanced in two ways. Firstly, constrained dictionaries were introduced to allow sample complexity reduction. Secondly, hierarchical atom search has been presented to reduce time complexity. The performance of the resulting network was assessed using realistic channel data with hardware impairments: it showed that our network outperforms classical approaches such as the LS estimator or the MMSE estimator with substituted covariance matrix, but also the classical mpNet network. This combined approach is really promising as it allows to enhance the estimation performance while reducing the overall complexity: this follows the efficiency philosophy of model-based deep learning~\cite{Shlezinger2020}. Future work could for example consider the adaptation of the proposed techniques to MU-MIMO-OFDM scenarios.


\bibliographystyle{IEEEtran}
\bibliography{biblio_paper.bib}

\begin{thebibliography}{10}
\providecommand{\url}[1]{#1}
\csname url@samestyle\endcsname
\providecommand{\newblock}{\relax}
\providecommand{\bibinfo}[2]{#2}
\providecommand{\BIBentrySTDinterwordspacing}{\spaceskip=0pt\relax}
\providecommand{\BIBentryALTinterwordstretchfactor}{4}
\providecommand{\BIBentryALTinterwordspacing}{\spaceskip=\fontdimen2\font plus
\BIBentryALTinterwordstretchfactor\fontdimen3\font minus
  \fontdimen4\font\relax}
\providecommand{\BIBforeignlanguage}[2]{{%
\expandafter\ifx\csname l@#1\endcsname\relax
\typeout{** WARNING: IEEEtran.bst: No hyphenation pattern has been}%
\typeout{** loaded for the language `#1'. Using the pattern for}%
\typeout{** the default language instead.}%
\else
\language=\csname l@#1\endcsname
\fi
#2}}
\providecommand{\BIBdecl}{\relax}
\BIBdecl

\bibitem{OShea2017}
T.~O’Shea and J.~Hoydis, ``An introduction to deep learning for the physical
  layer,'' \emph{IEEE Trans. Cogn. Commun. Netw.}, vol.~3, no.~4, pp. 563--575,
  2017.

\bibitem{Wang2017}
T.~Wang, C.-K. Wen, H.~Wang, F.~Gao, T.~Jiang, and S.~Jin, ``Deep learning for
  wireless physical layer: Opportunities and challenges,'' \emph{China
  Communications}, vol.~14, no.~11, pp. 92--111, 2017.

\bibitem{Lemagoarou2022}
L.~Le~Magoarou, T.~Yassine, S.~Paquelet, and M.~Crussi{\`e}re, ``Deep learning
  for location based beamforming with {NLOS} channels,'' in \emph{IEEE Int.
  Conf. Acoust., Speech, Signal Process. (ICASSP)}, 2022, pp. 8812--8816.

\bibitem{Alkhateeb2018}
A.~Alkhateeb, S.~Alex, P.~Varkey, Y.~Li, Q.~Qu, and D.~Tujkovic, ``Deep
  learning coordinated beamforming for highly-mobile millimeter wave systems,''
  \emph{IEEE Access}, vol.~6, pp. 37\,328--37\,348, 2018.

\bibitem{Yassine2022b}
T.~Yassine, L.~{Le Magoarou}, S.~Paquelet, and M.~Crussi{\`e}re, ``Leveraging
  triplet loss and nonlinear dimensionality reduction for on-the-fly channel
  charting,'' in \emph{IEEE Int. Workshop on Signal Process. Advances in
  Wireless Commun. (SPAWC)}, 2022.

\bibitem{Studer2018}
C.~Studer, S.~Medjkouh, E.~Gönültaş, T.~Goldstein, and O.~Tirkkonen,
  ``Channel charting: Locating users within the radio environment using channel
  state information,'' \emph{IEEE Access}, vol.~6, pp. 47\,682--47\,698, 2018.

\bibitem{Ferrand2021}
P.~Ferrand, A.~Decurninge, L.~G. Ordoñez, and M.~Guillaud, ``Triplet-based
  wireless channel charting: Architecture and experiments,'' \emph{IEEE J. Sel.
  Areas Commun.}, vol.~39, pp. 2361--2373, 2021.

\bibitem{LeMagoarou2021a}
L.~Le~Magoarou, ``Similarity-based prediction for channel mapping and user
  positioning,'' \emph{IEEE Commun. Lett.}, vol.~25, pp. 1578--1582, 2021.

\bibitem{Alrabeiah2019}
M.~Alrabeiah and A.~Alkhateeb, ``Deep learning for {TDD} and {FDD} {M}assive
  {MIMO}: {M}apping channels in space and frequency,'' in \emph{2019 53rd
  Asilomar Conf. Signals, Syst., Comput.}, 2019, pp. 1465--1470.

\bibitem{Shlezinger2020}
N.~Shlezinger, J.~Whang, Y.~C. Eldar, and A.~G. Dimakis, ``Model-based deep
  learning,'' \emph{arXiv preprint arXiv:2012.08405}, Dec. 2020.

\bibitem{Monga2021}
V.~Monga, Y.~Li, and Y.~C. Eldar, ``Algorithm unrolling: Interpretable,
  efficient deep learning for signal and image processing,'' \emph{IEEE Signal
  Process. Mag.}, vol.~38, no.~2, pp. 18--44, 2021.

\bibitem{Hershey2014}
J.~R. Hershey, J.~L. Roux, and F.~Weninger, ``Deep unfolding: Model-based
  inspiration of novel deep architectures,'' \emph{arXiv preprint
  arXiv:1409.2574}, 2014.

\bibitem{Balatsoukas2019}
A.~Balatsoukas-Stimming and C.~Studer, ``Deep unfolding for communications
  systems: A survey and some new directions,'' in \emph{2019 {IEEE} Int.
  Workshop on Signal Process. Syst. (SiPS)}, 2019, pp. 266--271.

\bibitem{Kang2008}
T.~Kang and R.~A. Iltis, ``Matching pursuits channel estimation for an
  underwater acoustic {OFDM} modem,'' in \emph{2008 IEEE Int. Conf. Acoust.,
  Speech, Signal Process. (ICASSP)}, 2008, pp. 5296--5299.

\bibitem{Karabulut2004}
G.~Z. Karabulut and A.~Yongacoglu, ``Sparse channel estimation using orthogonal
  matching pursuit algorithm,'' in \emph{IEEE 60th {IEEE} Veh. Technol. Mag.
  (VTC Fall)}, vol.~6, 2004, pp. 3880--3884 Vol. 6.

\bibitem{Khojastepour2009}
M.~A. Khojastepour, K.~Gomadam, and X.~Wang, ``Pilot-assisted channel
  estimation for {MIMO} {OFDM} systems using theory of sparse signal
  recovery,'' in \emph{2009 IEEE Int. Conf. Acoust., Speech, Signal Process.
  (ICASSP)}, 2009, pp. 2693--2696.

\bibitem{Yassine2022}
T.~Yassine and L.~Le~Magoarou, ``{mpNet}: variable depth unfolded neural
  network for massive {MIMO} channel estimation,'' \emph{IEEE Trans. Wireless
  Commun.}, vol.~21, no.~7, pp. 5703--5714, 2022.

\bibitem{Mallat1993}
S.~Mallat and Z.~Zhang, ``Matching pursuits with time-frequency dictionaries,''
  \emph{IEEE Trans. Signal Process.}, vol.~41, no.~12, pp. 3397--3415, 1993.

\bibitem{Kim1998}
D.~K. Kim, S.~H. Do, H.~B. Cho, H.~J. Chol, and K.~B. Kim, ``A new joint
  algorithm of symbol timing recovery and sampling clock adjustment for {OFDM}
  systems,'' \emph{IEEE Trans. Consum. Electron.}, vol.~44, no.~3, pp.
  1142--1149, 1998.

\bibitem{Speth1999}
M.~Speth, S.~A. Fechtel, G.~Fock, and H.~Meyr, ``Optimum receiver design for
  wireless broad-band systems using {OFDM}. {I},'' \emph{IEEE Trans. on
  Commun.}, vol.~47, no.~11, pp. 1668--1677, 1999.

\bibitem{Alkhateeb2019}
A.~Alkhateeb, ``{DeepMIMO}: A generic deep learning dataset for millimeter wave
  and massive {MIMO} applications,'' in \emph{Proc. of Inf. Theory and
  Applications Workshop (ITA)}, San Diego, CA, Feb 2019, pp. 1--8.

\bibitem{Kingma2014}
\BIBentryALTinterwordspacing
D.~P. Kingma and J.~Ba, ``Adam: A method for stochastic optimization,'' 2014.
  [Online]. Available: \url{https://arxiv.org/abs/1412.6980}
\BIBentrySTDinterwordspacing

\bibitem{Savaux2017}
V.~Savaux and Y.~Lou{\"e}t, ``{LMMSE channel estimation in {OFDM} context: a
  review},'' \emph{{IET Signal Process.}}, vol.~11, no.~2, pp. 123--134, 2017.

\end{thebibliography}

\end{document}